\documentclass{llncs}
\usepackage{llncsdoc}
\usepackage{graphicx}
\usepackage{subfigure}
\usepackage{makeidx}
\usepackage{hyperref}
\begin{document}

\newcounter{save}\setcounter{save}{\value{section}}
{\def\addtocontents#1#2{}%
\def\addcontentsline#1#2#3{}%
\def\markboth#1#2{}%
\title{An Adaptive XP-based approach to Agile Development}

\author{Gang Liao \and Lei Liu \and Lian Luo}

\institute{Computer Science and Engineering Department,
 Sichuan University Jinjiang College, 620860
 Penshan, China\\
 GreenHat1016@Gmail.com,cys19900611@gmail.com,mr.l172586418@gmail.com}

\maketitle

\begin{abstract}
Software design is gradually becoming open, distributed, pervasive, and
connected. It is a sad statistical fact that software projects are scientifically fragile
and tend to fail more than other engineering fields. Agile development is a philosophy.
And agile methods are processes that support the agile philosophy. XP places a strong
emphasis on technical practices in addition to the more common teamwork and structural practices.
In this paper, we elaborate how XP practices can be used to thinking, collaborating, releasing, planning, developing. And the state that make your team and organization more successful.

\end{abstract}

\section{Introduction}
Agile software development \cite{1} is a group of software development methods based on iterative and incremental development, where requirements and solutions evolve through collaboration between self-organizing, cross-functional teams. It promotes adaptive planning, evolutionary development and delivery, a time-boxed iterative approach, and encourages rapid and flexible response to change. It is a conceptual framework that promotes foreseen interactions throughout the development cycle.

Incremental software development methods have been traced back to 1957\cite{2}. In 1974, a paper by E. A. Edmonds introduced an adaptive software development  process\cite{3}. Concurrently and independently the same methods were developed and deployed by the New York Telephone Company{'}s Systems Development Center under the direction of Dan Gielan. In the early 1970s, Tom Gilb started publishing the concepts of Evolutionary Project Management (EVO), which has evolved into Competitive Engineering\cite{4}.

So-called lightweight software development methods evolved in the mid-1990s as a reaction against heavyweight methods, which were characterized by their critics as a heavily regulated, regimented, micromanaged, waterfall model of development. Proponents of lightweight methods (and now agile methods) contend that they are a return to development practices from early in the history of software development\cite{2}.Early implementations of lightweight methods include Scrum , Crystal Clear, Extreme Programming (XP) ,Adaptive Software Development, Feature Driven Development, and Dynamic Systems Development Method(DSDM) (1995). These are now typically referred to as agile methodologies, after the Agile Manifesto published in 2001\cite{4}.

They published the Manifesto for Agile Software Development \cite{1} to define the approach now known as agile software development. Some of the manifesto{'}s authors formed the Agile Alliance, a nonprofit organization that promotes software development according to the manifesto{'}s principles. Extreme programming (XP) is a software development methodology which is intended to improve software quality and responsiveness to changing customer requirements. As a type of agile software development \cite{5}.it advocates frequent "releases" in short development cycles (timeboxing), which is intended to improve productivity and introduce checkpoints where new customer requirements can be adopted.

In this article, we would exploit Extreme programming to analysis agile development what state is appropriate.

\section{Fail Statistic}
It{'}s a sad statistical fact that software projects are scientifically fragile and tend to fail. The intensiveness of the failure also varies from project to project. In a commercial software company, a failure will be related to the software consumer. The software did not meet with the consumers need, and the software release was later than scheduled (deadline violation) or had too many bugs. The following statistics are the ratio of result Fig.\ref{figc1} and failure reasons as follows:

\begin{figure}
\centering
\includegraphics[width=0.7\textwidth]{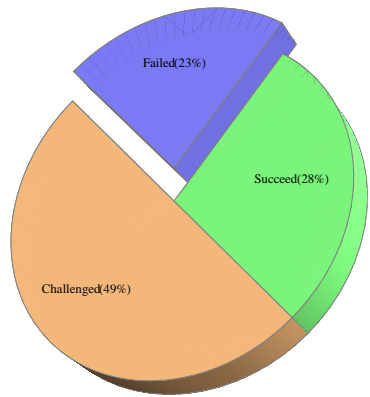}
\caption{the ratio of result.}
\label{figc1}
\end{figure}

\begin{tabular}{|l|c|}
\hline
\textbf{Project Challenged Factors} & \textbf{\% of Responses}\\
\hline
\hline
Lack of User Input & 12.8\%\\
\hline
Incomplete Requirements and Specifications & 12.3\%\\
\hline
Changing Requirements and Specifications & 11.8\%\\
\hline
Lack of Executive Support & 7.5\%\\
\hline
Technology Incompetence & 7.0\%\\
\hline
Lack of Resources & 6.4\%\\
\hline
Unrealistic Expectations & 5.9\%\\
\hline
Unclear Objectives & 5.3\%\\
\hline
Unrealistic Time Frames & 4.3\%\\
\hline
New Technology & 3.7\%\\
\hline
Other & 23.0\%\\
\hline
\end{tabular}\\\\

Some of the reasons are claims that were measured by researchers. I have tried to add my own judgment. It{'}s left to the reader. The software engineering field is much younger than the other engineering fields and that, in time, will get more stable. The field is young and therefore most of the field engineers and managers are also young. Young people have less experience and therefore tend to fail more.

As opposed to other engineering fields like civil engineering, the software engineering building blocks are much less tangible and therefore hard to measure and estimate \cite{6}. The competition in the software industry is harsh. The Time-To-Market (TTM) is crucial and the drive to meet harsh deadlines is enormous. This characteristic, along with other methodological anomalies like "Code first; think later" and "Plan to throw one away; you will, anyhow," makes competition harsh. The hard competition in the software industry causes not only the need to deliver ASAP, but also the requirement to catch as many potential customer eyes as possible. Firing in every direction causes disorganization, fast coding and projects that are not well planned.

Software development technologies change faster than other construction technologies \cite{6}.  Until recently, Microsoft was frequently bombarding the industry with new technologies. Rapid technology changes introduce liability for software manufactures. Developing life cycle methodology must be part of software project management. Nevertheless, it should not be forced into the R\&D environment. The software engineering field is relatively young , but still there are already well-known developing life cycle methodologists (Agile, Crystal, Spiral, Waterfall, etc.), successful stories and case studies.

The following Professor Brooks rule of thumb might seem radical \cite{7}, but being given no proportional time for planning and testing is indeed problematic: "1/3 of the schedule for design, 1/6 for coding, 1/4 for component testing, and 1/4 for system testing. "\cite{7}.  Tester to developer ratio: there is no rule of thumb that defines the number of QA engineers per software engineer. The reason for that is that it depends on many variables and more specifically on the characteristics of the software. There are several models that help with Tester to developer ratio (Fig.\ref{figc2}). According to a recent informal survey held at QAI{'}s 20th Annual Software Testing Conference in September of 2000:

\begin{figure}
\centering
\includegraphics[width=0.7\textwidth]{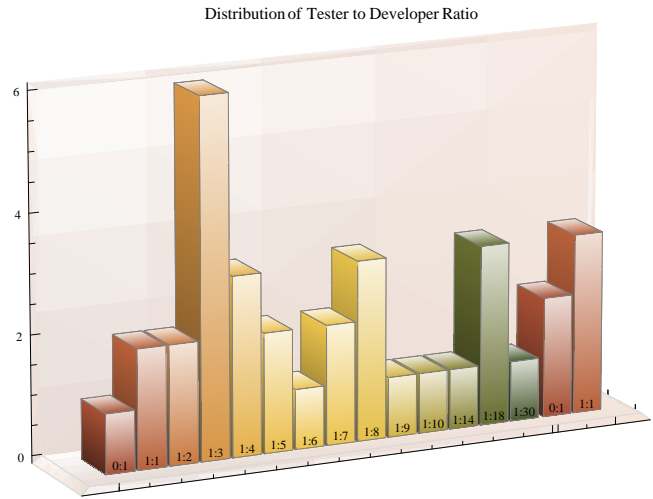}
\caption{Tester to developer ratio.}
\label{figc2}
\end{figure}

"＃Software development isn{'}t just a process of creating software; it{'}s also a process of learning how to create the software that is best suited for its purpose."\cite{6} The article describes some of the answers to the question, "Why software projects tend to fail?" I encourage the reader to keep reading on that topic for two main reasons:
\begin{itemize}
\item[.] Knowledge: knowing why software projects fail is a good start to preventing your own software project failure.
\item[.] Incomplete: the information in this article is incomplete; I consider it as a promo to keep reading.
\end{itemize}

\section{The XP Practices}
One of the most astonishing premises of XP \cite{8} is that you can eliminate requirements, design, and testing phases as well as the formal documents that go with them. This premise is so far off from the way we typically learn to develop software that many people dismiss it as a delusional fantasy. XP emphasizes face-to-face collaboration. This is so effective in eliminating communication
delays and misunderstandings that the team no longer needs distinct phases. This allows them to work on all activities every day〞with simultaneous phases〞as shown in Fig.\ref{figc3}.
\begin{figure}
\centering
\includegraphics[width=0.7\textwidth]{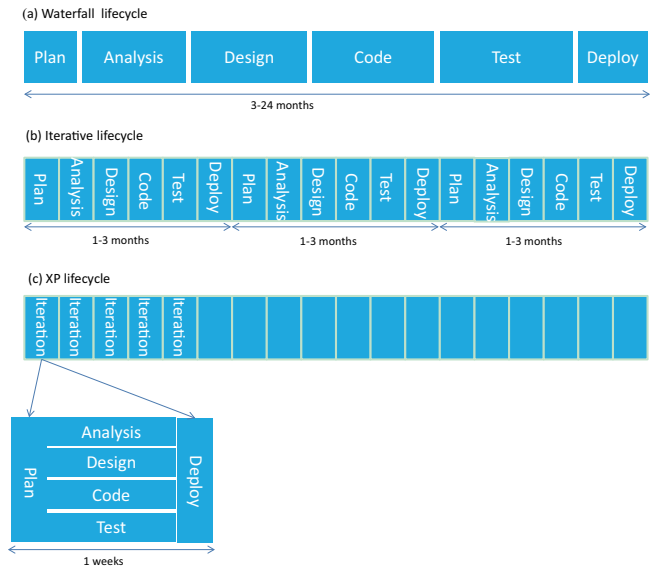}
\caption{(a) Waterfall lifecycle (b) Iterative lifecycle (c) XP lifecycle}
\label{figc3}
\end{figure}

Using simultaneous phases, an XP team produces deployable software every week. In each iteration, the team analyzes, designs, codes, tests, and deploys a subset of features. Although this approach doesn{'}t necessarily mean that the team is more productive, it does mean that the team gets feedback much more frequently(Fig.\ref{figc4}). As a result, the team can easily connect successes and failures to their underlying causes. The amount of unproven work is very small, which allows the team to correct some mistakes on the fly, as when coding reveals a design flaw, or when a customer review reveals that a user interface layout is confusing or ugly.

The tight feedback loop also allows XP teams to refine their plans quickly. It{'}s much easier for a customer to refine a feature idea if she can request it and start to explore a working prototype within a few days. The same principle applies for tests, design, and team policy. Any information you learn in one phase can change the way you think about the rest of the software. If you find a design defect during coding or testing, you can use that knowledge as you continue to analyze requirements and design the system in subsequent iterations.

XP teams perform nearly every software development activity simultaneously. Analysis, design, coding, testing, and even deployment occur with rapid frequency. That{'}s a lot to do simultaneously. XP does it by working in iterations: week-long increments of work. Every week, the team does a bit of release planning, a bit of design, a bit of coding, a bit of testing, and so forth. They work on stories: very small features, or parts of features, that have customer value. Every week, the team commits to delivering four to ten stories. Throughout the week, they work on all phases of development for each story. At the end of the week, they deploy their software for internal review.(In some cases, they deploy it to actual customers.) The following sections show how traditional phase-based activities correspond to XP iteration.

\subsection{Planning}
Every XP team includes several business experts,the on-site customers who are responsible for making
business decisions \cite{9}. The on-site customers point the project in the right direction by clarifying the project vision, creating stories, constructing a release plan, and managing risks. Programmers provide estimates
and suggestions, which are blended with customer priorities in a process called the planning game.
Together, the team strives to create small, frequent releases that maximize value.
The planning effort is most intense during the first few weeks of the project. During the remainder of
the project, customers continue to review and improve the vision and the release plan to account for
new opportunities and unexpected events.
In addition to the overall release plan \cite{10}, the team creates a detailed plan for the upcoming week at the
beginning of each iteration. The team touches base every day in a brief stand-up meeting, and its
informative workspace keeps everyone informed about the project status.

\begin{figure}
\centering
\includegraphics[width=0.6\textwidth]{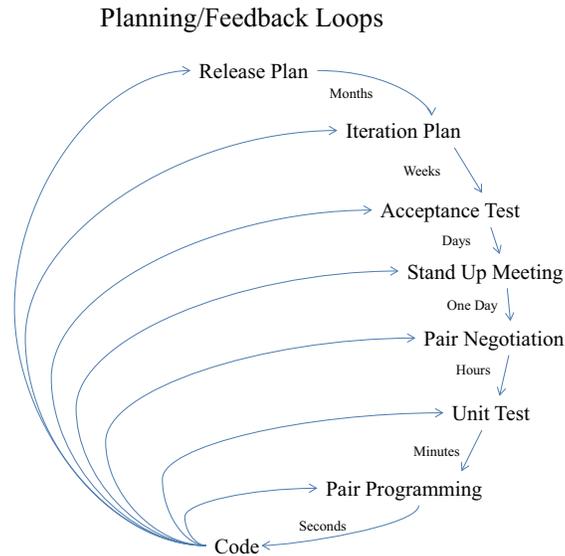}
\caption{High-level flow diagram of Feedback Loops}
\label{figc4}
\end{figure}

\subsection{Analysis}
Rather than using an upfront analysis phase to define requirements, on-site customers sit with the team
full-time. On-site customers may or may not be real customers depending on the type of project, but
they are the people best qualified to determine what the software should do.
On-site customers are responsible for figuring out the requirements for the software \cite{11}. To do so, they use
their own knowledge as customers combined with traditional requirements-gathering techniques.
When programmers need information, they simply ask. Customers are responsible for organizing their
work so they are ready when programmers ask for information. They figure out the general
requirements for a story before the programmers estimate it and the detailed requirements before the
programmers implement it.

Some requirements are tricky or difficult to understand. Customers formalize these requirements, with
the assistance of testers, by creating customer tests: detailed, automatically checked examples \cite{12}. Customers
and testers create the customer tests for a story around the same time that programmers implement the
story. To assist in communication, programmers use a ubiquitous language in their design and code.
The user interface (UI) look and feel doesn{'}t benefit from automated customer tests. For the UI,
customers work with the team to create sketches of the application screens. In some cases, customers
work alongside programmers as they use a UI builder to create a screen. Some teams include an
interaction designer who{'}s responsible for the application{'}s UI.

\subsection{Design and Coding}
XP uses incremental design and architecture to continuously create and improve the design in small
steps. This work is driven by test-driven development (TDD), an activity that inextricably weaves together
testing, coding, design, and architecture. To support this process, programmers work in pairs, which
increases the amount of brainpower brought to bear on each task and ensures that one person in each
pair always has time to think about larger design issues \cite{13}.

Programmers are also responsible for managing their development environment. They use a version
control system for configuration management and maintain their own automated build. Programmers
integrate their code every few hours and ensure that every integration is technically capable of
deployment.

To support this effort, programmers also maintain coding standards and share ownership of the code.
The team shares a joint aesthetic for the code, and everyone is expected to fix problems in the code
regardless of who wrote it.

\subsection{Testing}
XP includes a sophisticated suite of testing practices \cite{14}. Each member of the team〞programmers,
customers, and testers〞makes his own contribution to software quality. Well-functioning XP teams
produce only a handful of bugs per month in completed work.

Programmers provide the first line of defense with test-driven development. TDD produces automated
unit and integration tests. In some cases, programmers may also create end-to-end tests. These tests
help ensure that the software does what the programmers intended.
Likewise, customer tests help ensure that the programmers{'} intent matches customers{'} expectations.
Customers review work in progress to ensure that the UI works the way they expect. They also produce
examples for programmers to automate that provide examples of tricky business rules \cite{15}.
Finally, testers help the team understand whether their efforts are in fact producing high-quality code.
They use exploratory testing to look for surprises and gaps in the software. When the testers find a bug,
the team conducts root-cause analysis and considers how to improve their process to prevent similar
bugs from occuring in the future. Testers also explore the software{'}s nonfunctional characteristics, such
as performance and stability. Customers then use this information to decide whether to create additional
stories.

The team doesn{'}t perform any manual regression testing. TDD and customer testing leads to a
sophisticated suite of automated regression tests. When bugs are found, programmers create automated
tests to show that the bugs have been resolved. This suite is sufficient to prevent regressions. Every time
programmers integrate (once every few hours), they run the entire suite of regression tests to check if
anything has broken.

The team also supports their quality efforts through pair programming, energized work, and iteration
slack. These practices enhance the brainpower that each team member has available for creating high quality
software.

\subsection{Deployment}
XP teams keep their software ready to deploy at the end of any iteration\cite{16}. They deploy the software to internal stakeholders every week in preparation for the weekly iteration demo. Deployment to real
customers is scheduled according to business needs.

As long as the team is active, it maintains the software it has released. Depending on the organization,
the team may also support its own software (a batman is helpful in this case; see ※Iteration Planning§
in Chapter 8). In other cases, a separate support team may take over. Similarly, when the project ends,
the team may hand off maintenance duties to another team. In this case, the team creates
documentation and provides training as necessary during its last few weeks.

The following table shows how XP{'}s practices correspond to traditional phases. Remember that XP uses
iterations rather than phases; teams perform every one of these activities each week. Most are performed
every day.

\begin{tabular}{|l|c|c|c|c|c|}
\hline
\textbf{XP Practices} & \textbf{Planning} & \textbf{Analysis} & \textbf{Design$\&$Coding} & \textbf{Testing} &\textbf{Deployment}\\
\hline
\hline
\textbf{Thinking} & & & & & \\
\hline
Pair Programming & & & $\surd$ & $\surd$ & \\
\hline
Energized Work & $\surd$ & $\surd$ & $\surd$ & $\surd$ &$\surd$ \\
\hline
Informative Workspace & $\surd$ &  &  &  & \\
\hline
Root-Cause Analysis & $\surd$ &  &  & $\surd$ & \\
\hline
Retrospective & $\surd$ &  &  & $\surd$ & \\
\hline
\textbf{Collaborating} &  & &  & &\\
\hline
Trust & $\surd$ & $\surd$ & $\surd$ & $\surd$ &$\surd$ \\
\hline
Sit Together & $\surd$ & $\surd$ & $\surd$ & $\surd$ & \\
\hline
Real Customer Involvement &  & $\surd$ &  &  & \\
\hline
Ubiquitous Language &  & $\surd$ &  & & \\
\hline
Stand-Up Meetings &$\surd$  &  &  & & \\
\hline
Coding Standards &  &  &$\surd$  & & \\
\hline
Iteration Demo &  &  &  & &$\surd$ \\
\hline
Reporting &$\surd$  &$\surd$  &$\surd$  &$\surd$ &$\surd$ \\
\hline
\textbf{Releasing} &  & & & &\\
\hline
"Done Done" &  &  &$\surd$  & &$\surd$ \\
\hline
No Bugs &  & &$\surd$  &$\surd$ & \\
\hline
Version Control &  & &$\surd$  &$\surd$ & \\
\hline
Ten-Minute Build &  & &$\surd$  & &$\surd$ \\
\hline
Continuous Integration &  &  &$\surd$  & &$\surd$ \\
\hline
Collective Code Ownership  &  & &$\surd$  & & \\
\hline
Documents &  &  & & &$\surd$ \\
\hline
\textbf{Planning} & & & & & \\
\hline
Vision &$\surd$  &$\surd$ & & & \\
\hline
Release Planning &$\surd$  &$\surd$  &  & & \\
\hline
The Planning Game &$\surd$  &$\surd$  &  & & \\
\hline
Risk Management &$\surd$  & & & &\\
\hline
Iteration Planning &$\surd$  &  &$\surd$  & & \\
\hline
Customer Tests &  &$\surd$  &$\surd$  & & \\
\hline
Text-Driven Development &  &$\surd$  &$\surd$  & & \\
\hline
Refactoring &  &  &$\surd$  & & \\
\hline
Simple Design &  &  &$\surd$  & & \\
\hline
Incremental Architecture &  &  &$\surd$  & & \\
\hline
Spike Design &  &  &$\surd$  & & \\
\hline
Performance Optimization &  &  &$\surd$  & & \\
\hline
Exploratory &  &  &  & &$\surd$ \\
\hline
\end{tabular}\\\\

\section{Adopting XP}
Before adopting XP, you need to decide whether it{'}s appropriate for your situation \cite{17} \cite{18}. XP{'}s applicability is based on organizations and people, not types of projects.
XP{'}s applicability has far more to do with your organization and the people involved than with the type of project you{'}re working on. Similarly, if you want to practice XP, do everything you can to meet the following prerequisites and recommendations. This is a lot more effective than working around limitations.

\subsection{Management Support}
 It{'}s very difficult to use XP in the face of opposition from management. Active support is best. Active support is best. To practice XP as described,you will need the following:
 \begin{itemize}
\item[.] A common workspace with pairing stations.
\item[.] Team members solely allocated to the XP project.
\item[.] A product manager, on-site customers, and integrated testers.
\end{itemize}

You will often need management's help to get the previous three items. In addition, the more
management provides the following things, the better:
\begin{itemize}
\item[.] Team authority over the entire development process, including builds, database schema, and
version control
\item[.] Compensation and review practices that are compatible with team-based effort.
\item[.] Acceptance of new ways of demonstrating progress and showing results.
\item[.] Patience with lowered productivity while the team learns
\end{itemize}

\subsection{Team Agreement}
Just as important as management support is the team{'}s agreement to use XP \cite{18}. If team members don{'}t
want to use XP, it{'}s not likely to work. XP assumes good faith on the part of team members〞there's no
way to force the process on somebody who{'}s resisting it.
It{'}s never a good idea to force someone to practice XP against his will. In the best case, he{'}ll find some way to leave the team, quitting if necessary. In the worst case, he{'}ll remain on the team and silently sabotage your efforts \cite{19}.
Reluctant skeptics are OK. If somebody says, ※I don{'}t want to practice XP, but I see that the rest of you
do, so I{'}ll give it a fair chance for a few months,§ that{'}s fine. She may end up liking it. If not, after a few months have gone by, you{'}ll have a better idea of what you can do to meet the whole team{'}s needs.If only one or two people refuse to use XP, and they{'}re interested in working on another project, let them transfer so the rest of the team can use XP. If no such project is available, or if a significant portion of the team is against using XP, don{'}t use it.

\subsection{The Right Team Size}
For teams new to XP, however,I recommend 4 to 6 programmers and no more than 12 people on the team.
I also recommend having an even number of programmers so that everyone can pair
program. If you have ongoing support needs, add one more programmer for a total of five or seven so that the team can have a batman.
Teams with fewer than four programmers are less likely to have the intellectual diversity they need.
They{'}ll also have trouble using pair programming, an important support mechanism in XP. Large teams
face coordination challenges. Although experienced teams can handle those challenges smoothly, a new
XP team will struggle.

\subsection{A Brand-New Codebase}
Easily changed code is vital to XP \cite{20}. If your code is cumbersome to change, you{'}ll have difficulty with XP{'}s technical practices, and that difficulty will spill over into XP{'}s planning practices.
XP teams put a lot of effort into keeping their code clean and easy to change. If you have a brand-new
codebase, this is easy to do. If you have to work with existing code, you can still practice XP, but it will
be more difficult. Even well-maintained code is unlikely to have the simple design and suite of
automated unit tests that XP requires (and produces). New XP teams often experience an epiphany
between the second and fourth months. ※This is the best code I{'}ve ever worked with!§ they say, and
start to see the power of XP.

To understand and appreciate XP{'}s technical practices fully, you need to experience the practices
meshing together to give you complete confidence in your code, tests, and build. You need to feel the
delight of making big improvements with small changes. You{'}re unlikely to have that experience when
working with existing code. If you can, leave preexisting code to experienced XP teams.

\subsection{An Experienced Programmer-Coach}
Some people are natural leaders. They{'}re decisive, but appreciate others{'} views; competent, but
respectful of others{'} abilities. Team members respect and trust them. You can recognize a leader by her
influence〞regardless of her title, people turn to a leader for advice.

XP relies on self-organizing teams. This kind of team doesn't have a predefined hierarchy; instead, the
team decides for itself who is in charge of what. These roles are usually informal. In fact, in a mature
XP team, there is no one leader. Team members seamlessly defer leadership responsibilities from one
person to the next, moment to moment, depending on the task at hand and the expertise of those
involved.
When your team first forms, though, it won{'}t work together so easily. Somebody will need to help the
team remember to follow the XP practices consistently and rigorously. This is particularly important for
programmers, who have the most difficult practices to learn.

In other words, your team needs a coach. The best coaches are natural leaders〞people who remind
others to do the right thing by virtue of who they are rather than the orders they give. Your coach also
needs to be an experienced programmer so she can help the team with XP{'}s technical practices.

\section{Assess Your Agility}
Suppose you{'}ve been using XP for a few months. How can you tell if you{'}re doing it properly? The
ultimate measure is the success of your project, but you may wish to review and assess your approach
to XP as well.

To help you do this, I{'}ve created a quiz that focuses on five important aspects of agile development. It
explores results rather than specific practices, so you can score well even after customizing XP to your
situation. If you aren{'}t using XP at all, you can also use this quiz to assess your current approach.
This quiz assesses typical sources of risk. Your goal should be to achieve the maximum score in each
category〞which is well within the grasp of experienced XP teams. Any score less than the maximum
indicates risk, and an opportunity for improvement.

To take the quiz, answer the following questions and enter your scores on a photocopy of the blank
radar diagram (Fig.\ref{figc6}). Don{'}t give partial credit for any question, and if you aren{'}t sure of the answer, give yourself zero points. The result should look something like Fig.\ref{figc5} The score of the lowest spoke identifies your risk, as follows:

\begin{itemize}
\item[.] 75 points or less: immediate improvement required
\item[.] 75 to 96 points: improvement necessary
\item[.] 97, 98, or 99: improvement possible
\item[.] 100: no further improvement needed
\end{itemize}

\begin{figure}
\centering
\includegraphics[width=0.8\textwidth]{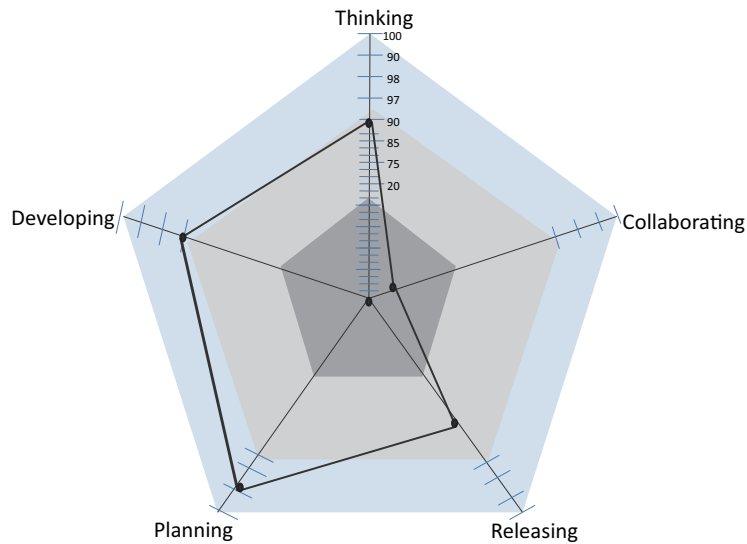}
\caption{The Result of Quizzing.}
\label{figc5}
\end{figure}

\begin{figure}
\centering
\includegraphics[width=0.8\textwidth]{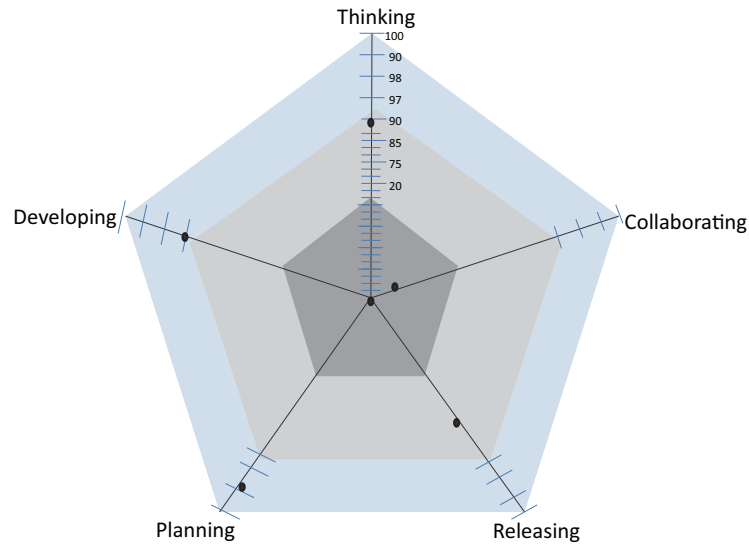}
\caption{Self-Assessment Quiz.}
\label{figc6}
\end{figure}

\section{Conclusion}
It{'}s a fact of life: change makes people uncomfortable. XP is probably a big change for your team. If you
previously used a rigid, document-centric process, XP will seem loose and informal. If you previously
had no process, XP will seem strict and disciplined. Either way, expect team members and stakeholders
to be uncomfortable. This discomfort can extend into the larger organization.
Discomfort and a feeling of chaos is normal for any team undergoing change, but that doesn{'}t make it
less challenging. Expect the chaotic feeling to continue for at least two months. Give yourselves four to
nine months to feel truly comfortable with your new process. If you{'}re adopting XP incrementally, it
will take longer.

To survive the transformation, you need to know why you are making this change. What benefits does
it provide to the organization? To the team? Most importantly, what benefits does it provide to each
individual? As you struggle with the chaos of change, remember these benefits.
A supportive work environment is also important. Team members are likely to experience defense
reactions to the lack of familiar structure. Expect mood swings and erratic behavior. Some team
members may lash out or refuse to cooperate. Acknowledge the discomfort people are experiencing,
and help team members find constructive outlets for their frustration.

Your stakeholders may be uncomfortable with your team{'}s new approach to planning and reporting
progress. Managers and executives may see the team{'}s initial chaos as a sign that XP won{'}t work. To
help everyone feel more comfortable, consider giving them this pledge:
Our pledge to users, management, and other stakeholders.
We promise to:

\begin{itemize}
\item[.] Make steady progress
\item[.] Finish the features that you consider most valuable first
\item[.] Show you working software that reflects our progess every week, on (day of week) at
(time) in (location)
\item[.] Be honest and open with you about our successes, challenges, and what we can reasonably
provide
\end{itemize}

We believe that XP-Based can be properly modeled using techniques form software engineering and this article is just an early trail of using XP to improve Improve efficiency and success.

\end{document}